\documentclass[seceq]{ptptex}

\usepackage{graphicx}

\def\be{\begin{equation}}
\def\ee{\end{equation}}
\def\bea{\begin{eqnarray}}
\def\eea{\end{eqnarray}}

\title{
Thermodynamic systems as  bosonic strings%
}

\author{
Alejandro \textsc{V\'azquez}$^1$, Hernando \textsc{Quevedo}$^2$, and Alberto \textsc{S\'anchez}$^3$
}

\inst{
$^1$ Facultad de Ciencias, 
Universidad Aut\'onoma de Morelos, 
Av. Universidad 1001,   
 Cuernavaca, MO 62210, Mexico\\
$^2$ Dipartimento di Fisica, Universit\`a di Roma, Piazzale Aldo Moro 5, I-00185 Roma;
ICRANet, Piazzale della Repubblica 10, I-65122 Pescara, Italy\\
            $^3$ Instituto de Ciencias Nucleares, 
Universidad Nacional Aut\'onoma de M\'exico  
AP 70543, M\'exico, DF 04510, Mexico
}

\abst{
We apply variational principles in the context 
of geometrothermodynamics. The thermodynamic phase space ${\cal T}$ and the
space of equilibrium states ${\cal E}$ 
turn out to be described by Riemannian metrics
which are invariant with respect to Legendre transformations
and satisfy the differential equations following from 
the variation of a Nambu-Goto-like action. 
This implies that
the volume element of ${\cal E}$ is an extremal and that 
${\cal E}$ and ${\cal T}$ are related by an embedding harmonic map.  
  We explore the
physical meaning of geodesic curves in ${\cal E}$ as describing 
quasi-static processes that connect
different equilibrium states.
We present a Legendre invariant metric which is flat (curved) in the case of an ideal (van 
der Waals) gas and satisfies Nambu-Goto equations. The method is used to derive some
new solutions which could represent particular thermodynamic 
systems. 
}

\begin{document}

\maketitle

\section{Introduction}
\label{sec:int}

The geometry of thermodynamics has been the subject of moderate research
since the original works by Gibbs \cite{gibbs} and
Caratheodory \cite{car}. Results have been achieved in 
two different approaches.  
The first one consists in introducing metric structures on the space
of thermodynamic equilibrium  states ${\cal E}$, whereas the second 
group uses the contact structure of the thermodynamic phase space 
${\cal T}$. 
Weinhold \cite{wei1} introduced on ${\cal E}$ a metric, 
defined as  the Hessian  of the internal energy, which is related by a conformal transformation 
to the Hessian of the entropy (Ruppeiner's metric).   
This approach has found 
applications also in the context  of  the renormalization
group \cite{hung1}, the identification of states to a specific 
phase \cite{hung2}, the thermodynamic uncertainty relations \cite{ital},
and black hole thermodynamics (see \cite{quev08grg,quevaz} and references cited there).
The second approach, proposed by Hermann \cite{her},
uses the natural contact structure of the phase space ${\cal T}$.
Extensive and intensive thermodynamic
variables are taken together with the thermodynamic potential to
constitute coordinates on ${\cal T}$.  
A special subspace of ${\cal T}$ is the space of equilibrium states
${\cal E}$.

Geometrothermodynamics (GTD) \cite{quezar03,quev07} was recently developed as 
a formalism that  
unifies the contact structure on ${\cal T}$ with the metric structure 
on ${\cal E}$ in a consistent manner, by considering only 
Legendre invariant metrics on ${\cal T}$. 
This last property is important to guarantee that the properties
of a system do not depend on the thermodynamic potential 
used for its description. One simple metric \cite{quev07} was used in GTD in order
to reproduce geometrically the noncritical and critical behavior of the
ideal and van der Waals gas \cite{callen}, respectively, indicating that 
thermodynamic
curvature can be used as a measure of thermodynamic interaction. This
result has been corroborated  in the case of black holes \cite{aqs08,qs08,qs08btz}.

In the present work we explore an additional aspect of GTD. The thermodynamic 
metrics used so far in GTD, have been derived by using only the  
Legendre invariance condition. Now we ask the question whether those metrics 
can be derived as solutions of a certain set of differential equations, as 
it is usual in field theories. 
We will see that this task is realizable. In fact, it turns out that the 
map $\varphi: {\cal E} \rightarrow {\cal T}$ can be considered 
as a harmonic map, if the thermodynamic variables satisfy the 
differential equations which follow from the variation of a Polyakov-like
action. This result confers 
thermodynamics a geometric structure that resembles that of the bosonic 
string theory. In fact, we will show that thermodynamic systems can be interpreted as ``strings" 
embedded in a higher dimensional, curved phase space.

This paper is organized as follows. In Section \ref{sec:gtd} we review the 
fundamentals of GTD. In Section  \ref{sec:hm} we show that the embedding map 
$\varphi$ can be considered as a harmonic map with a naturally induced 
Polyakov-like action from which a set of differential equations can be derived. 
 Section \ref{sec:geo} contains an 
analysis of the thermodynamic length, the variation of which leads to the 
geodesic equations for the metric of ${\cal E}$. In Section \ref{sec:appl} we present 
the most general Legendre invariant metric we have found and use it to derive 
the geometric structure of 
thermodynamic systems with an arbitrary
 finite number of different species. As concrete examples
we investigate the geometry of the ideal gas, the van der Waals gas, and derive 
a few new fundamental equations which are compatible with the geometric structures of
GTD. 
Finally, Section \ref{sec:con} is devoted to a discussion of our results. 

\section{Geometrothermodynamics}
\label{sec:gtd}

Consider the $(2n+1)$-dimensional space ${\cal T}$ 
coordinatized by the set $Z^A=\{\Phi, E^a, I^a\}$, with the notation
$A=0,...,2n$ so that $\Phi=Z^0$, $E^a = Z^a$, $ a=1,...,n$, and $I^a = Z^{n+a}$.
Here $\Phi$ represents the thermodynamic potential, $E^a$ are the extensive variables
and $I^a$ the intensive variables. Consider 
the Gibbs 1-form \cite{her}
\be
 \Theta = d\Phi - \delta_{ab} I^a d E^b \ ,\quad \delta_{ab}={\rm diag} (1,1,...,1)
\label{gibbs}
\ee
where summation over repeated indices is assumed. The pair
$({\mathcal T}, \Theta)$ is called a contact manifold \cite{her},
 if ${\mathcal T}$ is differentiable and $\Theta$ satisfies the condition 
$\Theta \wedge (d\Theta)^n \neq 0$. It can be shown \cite{her} that if there exists a second differential form $\tilde \Theta$ which satisfies the 
condition $\tilde \Theta \wedge (d\tilde \Theta)^n \neq 0$ on ${\cal T}$, then $\Theta$ and $\tilde\Theta$ must be related by
a Legendre transformation \cite{arnold}, 
$
\{Z^A\}\rightarrow \{\widetilde{Z}^A\}=\{\tilde \Phi, \tilde E ^a, \tilde I ^ a\},
$
\be
 \Phi = \tilde \Phi - \delta_{kl} \tilde E ^k \tilde I ^l \ ,\quad
 E^i = - \tilde I ^ {i}, \ \  
E^j = \tilde E ^j,\quad   
 I^{i} = \tilde E ^ i , \ \
 I^j = \tilde I ^j \ ,
 \label{leg}
\ee
where $i\cup j$ is any disjoint decomposition of the set of indices $\{1,...,n\}$,
and $k,l= 1,...,i$. This implies that the contact structure of ${\cal T}$ is invariant
with respect to Legendre transformations. 
Consider, in addition, a nondegenerate metric $G$ on ${\cal T}$. 
We define the {\it thermodynamic phase space} 
as the triplet $({\cal T},\Theta, G)$ such that $\Theta$ defines a contact structure on ${\cal T}$ and
$G$ is a Legendre invariant metric
on ${\cal T}$. A straightforward computation shows that the 
flat Euclidean metric is not Legendre invariant. For 
simplicity, the phase space 
will be denoted  by ${\cal T}$. 
The {\it space of equilibrium states} is an $n-$dimensional Riemannian manifold 
$({\cal E},g)$, where ${\cal E} \subset {\cal T}$ is defined by a 
smooth map $ \varphi :   {\mathcal E}   \rightarrow {\mathcal T}$,
satisfying the conditions $\varphi^*(\Theta) =0$ and $g=\varphi^*(G)$,
where $\varphi^*$ is the pullback of $\varphi$. The smoothness of the map
$\varphi$ guarantees that $g$ is a well-defined, nondegenerate metric on ${\cal E}$.
For the sake of concreteness we choose $E^a$ as the coordinates of ${\cal E}$.
Then,  
$ \varphi :  \{E^a\} \mapsto \{\Phi(E^a), E^a, I^a(E^a)\}$, and the 
condition $\varphi^*(\Theta)=0$ yields the first law of thermodynamics and the
condition for thermodynamic equilibrium:
\be
d\Phi = \delta_{ab} I^a d E^b = I_b d E^b   \ , \qquad \frac{\partial\Phi}{\partial E^a} = 
\delta_{ab} I^b = I_a \ .
\label{equil}
\ee
The explicit form of $\varphi$ implies that 
the function $\Phi(E^a)$ must be given. In thermodynamics $\Phi(E^a)$ is known
as the fundamental equation from which all the equations of state
 of the system can be derived. It also satisfies the 
second law of thermodynamics which is equivalent
to the condition \cite{callen} 
\be
\frac{\partial^2 \Phi}{\partial E ^a \partial E ^b} \geq 0 \ .
\label{slaw}
\ee
In addition, the degree $\beta$ of homogeneity of the thermodynamic 
potential, i. e., $\Phi(\lambda E^a) = \lambda^\beta\Phi(E^a)$, with constant
$\lambda$ and $\beta$, appears then in Euler's identity: \cite{quev07}
\be
\label{euler}
\beta \Phi = E^aI_a = E^a\frac{\partial \Phi}{\partial E^a}\ .
\ee

The metric defined by $g=\varphi^*(G)$ represents
the {\it thermodynamic metric} of ${\cal E}$ whose components are  
explicitly given as
\be 
g_{ab} = \frac{\partial Z^A}{\partial E^a} \frac{\partial Z^B}{\partial E^b} G_{AB} 
= Z^A_{,a} Z^B_{,b} G_{AB} \ .
\label{gdown}
\ee
It is worth noticing that the application of a Legendre transformation 
on $G$ corresponds to a coordinate transformation of $g$. Indeed, if we denote
by $\{\tilde Z^A\}$ the Legendre transformed coordinates in ${\cal T}$, then
the transformed metric $ 
\tilde G_{AB} = \frac{\partial Z^C}{\partial \tilde Z^A}  
\frac{\partial Z^D}{\partial \tilde Z^B} G_{CD}
$
 induces on ${\cal E}$ the metric
$
\tilde g_{ab} =      
\frac{\partial \tilde Z^A}{\partial \tilde E^a}\frac{\partial \tilde Z^B}
{\partial \tilde E^b} \tilde G_{AB}
$.
It is then easy to see that these metrics are related
by the transformation law
\be
\tilde g_{ab} =  \frac{\partial E^c}{\partial \tilde E^a}
\frac{\partial E^d}{\partial \tilde E^b} g_{cd}\ .
\ee

\section{The harmonic map}
\label{sec:hm}
Consider the phase space with metric $G$ and coordinates $Z^A$,
and suppose that an arbitrary 
nondegenerate metric $h$ is given in ${\cal E}$.  
The smooth map $\varphi: {\cal E} \rightarrow {\cal T}$, or in coordinates
$\varphi: \{E^a\} \mapsto \{Z^A\}$,  is called a {\it harmonic map},  
if the coordinates $Z^A$ satisfy the equations following from the variation of
the action \cite{misner}
\be
I_h = \int_{\cal E} d^n E \, \sqrt{|h|}\,\, h^{ab}Z^A_{,a} Z^B_{,b} G_{AB} \ .
\label{hmaction}
\ee
Here $|h|=|{\rm det}(h_{ab})|$.
The variation of $I_h$ with respect to $Z^A$  leads to
\be
\frac{\delta I_h}{\delta Z^A}=0 \Leftrightarrow 
 {\cal D}_h  Z^A :=
\frac{1}{\sqrt{|h|}}\left(\sqrt{|h|}\, h^{ab}Z^A_{,a}\right)_{,b} + 
\Gamma^A_{\ BC} Z^B_{,b}Z^C_{,c} h^{bc}  = 0 \,
\ee
where $\Gamma^A_{\ BC}$ are the Christoffel symbols associated to 
the metric $G_{AB}$. 
For given metrics $G$ and $h$, this is a set of $2n+1$ second--order, 
partial differential equations for the $2n+1$ thermodynamic variables
$Z^A$. This set of equations must be treated together with the equation 
for the metric $h$, i.e.,
\be
\frac{\delta I_h}{\delta h^{ab}}=0 \Leftrightarrow T_{ab}:=
g_{ab} -\frac{1}{2}h_{ab} h^{cd} g_{cd} = 0 \ , 
\ee
where $g_{ab}$ is the metric induced on ${\cal E}$ by the pullback $\varphi^*$ 
according to (\ref{gdown}). This is an algebraic constraint from which it is easy to derive the expression
\be
h^{ab}g_{ab}= 2\left(\frac{|g|}{|h|}\right)^{1/2}
\ ,
\label{hg}
\ee
where $|g|=|\det(g_{ab})|$.
This analysis is similar to the analysis performed in string theory for the 
bosonic string by using the Polyakov action. In fact, action (\ref{hmaction}) 
with $n=2$ and a flat ``background", $G_{AB}=\eta_{AB}$, is known as the Polyakov action \cite{pol}. 
By analogy with string theory, from the above description we can conclude 
that a thermodynamic system with $n$ degrees of freedom can be interpreted 
as an $n-$di\-men\-sional ``membrane" which ``propagates"
 on a curved background metric $G$.  The quotations marks emphasize the fact
that at this level we have no explicit timelike parameter for the description 
of the membrane. Neither have we a timelike coordinate in the set $Z^A$ 
so that the metric $G$ could be interpreted as a background spacetime where the
membrane propagates. Nevertheless, this analogy allows us to handle thermodynamics
as a field theory where the thermodynamic variables satisfy a set of second--order 
differential equations. 
As in string theory, there is an equivalent description in terms of a Nambu-Goto-like
action. Introducing the relationship (\ref{hg}) into the action (\ref{hmaction})
and using the expression (\ref{gdown}) for the induced metric, we obtain 
the action $
I_g = 2
\int_{\cal E} d^n E \, \sqrt{|g|}$ from which we derive the Nambu-Goto equations
\be
{\cal D} _g Z^A = \frac{1}{\sqrt{|g|}}\left(\sqrt{|g|}\,\, g^{ab}Z^A_{,a}\right)_{,b} + 
\Gamma^A_{\ BC} Z^B_{,b}Z^C_{,c} g^{bc} =0 \ .
\label{meqg}
\ee
Since the action $I_g$ is proportional to the 
volume element of the manifold ${\cal E}$, equations (\ref{meqg}) can be 
interpreted as stating that the volume element induced in ${\cal E}$
must be an extremal. An equivalent interpretation 
is that the submanifold ${\cal E}$ can be represented as an extremal hypersurface 
contained in ${\cal T}$. 
The Nambu-Goto equations (\ref{meqg}) can be investigated by using standard procedures
of string theory.

\section{Geodesics in the space of equilibrium states}
\label{sec:geo}

In the space of equilibrium states ${\cal E}$
 the line element
$
ds^2 = g_{ab} dE^a  dE^b \ 
$
can be considered as a measure for the distance between two points $t_1$ and 
$t_2$ 
with coordinates $E^a$ and $E^a + dE^a$, respectively. 
 Let us assume that the points $t_1$ and 
$t_2$  belong to the curve $\gamma(\tau)$. Then, we define the {\it thermodynamic length} $L$
as
\be
L = \int_{t_1}^{t_2} ds = \int_{t_1}^{t_2} \left( g_{ab} dE^a  dE^b \right)^{1/2}
= \int_{t_1}^{t_2} \left( g_{ab} \dot E^a  \dot E^b \right)^{1/2}d\tau\ ,
\label{geo}
\ee
where the dot represents differentiation with respect to $\tau$.
The condition that the thermodynamic length be an extremal $
\delta L = \delta  \int_{t_1}^{t_2} ds = 0$ leads to the equations
\be
\frac{d ^2E^a}{d\tau^2} + 
\Gamma^a_{\ bc} \frac{dE^b}{d\tau} \frac{dE^c}{d\tau} = 0 \ ,
\label{geo1}
\ee
where $\Gamma^a_{\ bc}$ are the Christoffel symbols of the thermodynamic metric $g_{ab}$.
These are the geodesic equations in the space ${\cal E}$ with affine parameter $\tau$.
The solutions to the geodesic equations depend on the explicit form of the thermodynamic 
metric $g$ which, in turn, depends on the fundamental equation $\Phi=\Phi(E^a)$. Therefore,
a particular thermodynamic system leads to a specific set of geodesic equations whose 
solutions depend on the properties of the system. Not all the solutions need to be 
physically realistic since, in principle, there could be geodesics that connect equilibrium
states that are not compatible with the laws of thermodynamics. 
Those geodesics which connect 
physically meaningful states will represent quasi-static thermodynamic processes. Therefore,
a quasi-static process can be seen as a dense succession of equilibrium states. This is  in 
agreement with the standard interpretation of quasi-static processes in ordinary thermodynamics \cite{callen}.
The affine parameter $\tau$ can be used to label each of the equilibrium states which are 
part of a geodesic. Because of its intrinsic freedom, 
the affine parameter can be chosen in such a way that it increases as the 
entropy of a quasi-static process increases. This opens the possibility of interpreting the
affine parameter as a ``time" parameter with a specific direction which coincides with the
direction of entropy increase.

\section{Applications}
\label{sec:appl}
To solve the differential equations of GTD
(\ref{meqg}) 
one must specify {\it a priori} a Legendre invariant metric $G$ for the phase space \cite{quev07}. The metric 
\be
\label{ginv2}
G=\left(d\Phi - I_a dE^a\right)^2 + \Lambda\, (E_a I_a)^{2k+1} d E^a d I^a \ ,
\ee
with $k$ being an integer and $\Lambda$ an arbitrary Legendre invariant function, is the most general metric we have found that is invariant with respect to arbitrary Legendre transformations.  
To determine the metric structure of ${\cal E}\subset {\cal T}$ we choose the  map 
$ \varphi :  \{E^a\} \mapsto \{\Phi(E^a), E^a, I^a(E^a)\}$ so that
the condition $g=\varphi^*(G)$ for (\ref{ginv2})
yields
\be
g= \Lambda\left(E_a\frac{\partial\Phi}{\partial E^a}\right)^{2k+1} 
\frac{\partial^2\Phi}{\partial E^b \partial E^c}\delta^{ab} dE^a dE^c
\ ,
\label{gdowninv}
\ee
where we have used the first law of thermodynamics and the equilibrium conditions
 as given in Eqs.(\ref{equil}). 
In ordinary thermodynamics, the most used representation is based upon
the internal energy $U$. The extensive variables are chosen  
as the entropy $S$, volume $V$, and the particle number of each species $N_m$. For
simplicity we fix the maximum number of species as $n-2$ so that 
$m=1,...,n-2$. The 
dual variables are denoted as temperature $T$, pressure $-P$, and chemical 
potentials $\mu_m$. The coordinates of ${\cal T}$ are then $Z^A = \{U,S, V, N_m,T,-P,\mu_m\}$, and the
fundamental Gibbs 1-form becomes $\Theta = dU - TdS + P dV - \sum_m \mu_m dN_m$. 
As for the Riemannian structure of ${\cal T}$, the most general Legendre invariant metric (\ref{ginv2}) 
becomes 
\bea
\label{ginv3}
G= & &   \left(dU - TdS + P dV - \sum_{m=1}^{n-2} \mu_m d N_m\right)^2 \nonumber \\
& +&  \Lambda \left[ (ST)^{2k+1} dS dT + (VP)^{2k+1} dV dP + 
\sum_{m=1}^{n-2} (N_m\mu_m)^{2k+1} d N_m d\mu_m\right] \ .
\eea
For the space of equilibrium states ${\cal E}$ we choose the extensive
variables $E^a=\{S,V,N_m\}$ with the embedding map $\varphi:\{E^a\}
\mapsto \{Z^A\}$. Then, the condition $\varphi^*(\Theta)=0$ generates the 
first law of thermodynamics and the equilibrium conditions 
\be
dU = TdS - P dV + \sum_{m=1}^{n-2} \mu_m d N_m \ ,
\ee
\be
T=\frac{\partial U}{\partial S}\ , \quad
P= - \frac{\partial U}{\partial V}\ , \quad
\mu_m = \frac{\partial U}{\partial N_m}\ ,
\label{condte}
\ee
respectively. 
Furthermore, the metric of ${\cal E}$ is determined by (\ref{gdowninv}) that 
becomes
\bea
\label{gdowninv1}
g  = && \Lambda \Bigg\{ 
\left(S\frac{\partial U}{\partial S}\right)^{\tilde k}
\frac{\partial^2 U}{\partial S ^2} d S^2
+ \left(V\frac{\partial U}{\partial V}\right)^{\tilde k} 
\frac{\partial^2 U}{\partial V ^2} d V^2 + 
\sum_{m=1}^{n-2}
\left(N_m\frac{\partial U}{\partial N_m}\right)^{\tilde k} 
\frac{\partial^2 U}{\partial N_m ^2} dN_m^2
\protect\nonumber \\
& & +   \left[ \left(S\frac{\partial U}{\partial S}\right)^{\tilde k}
+\left(V\frac{\partial U}{\partial V}\right)^{\tilde k}\right]
\frac{\partial^2 U}{\partial S \partial V} d S d V  \nonumber \\
&& + 
\left[\left(S\frac{\partial U}{\partial S}\right)^{\tilde k} +
\sum_{m=1}^{n-2}
\left(N_m\frac{\partial U}{\partial N_m}\right)^{\tilde k}\right]
\frac{\partial^2 U}{\partial S \partial N_m} d S d N_m  \nonumber \\
&& + 
\left[\left(V\frac{\partial U}{\partial V}\right)^{\tilde k} +
\sum_{m=1}^{n-2}
\left(N_m\frac{\partial U}{\partial N_m}\right)^{\tilde k}\right]
\frac{\partial^2 U}{\partial V\partial N_m} d V d N_m  \Bigg\} \ ,
\eea
where $\tilde k =2k+1$. 
This is the most general metric corresponding to 
a multicomponent system with $n-2$ different species. This metric turns out to be useful 
to describe chemical reactions which can 
then be classified in accordance to the geometric properties of the manifold 
$({\cal E},g)$. This result will be presented elsewhere. 

\subsection{The ideal gas}
\label{sec:ig}
Consider a monocomponent ideal gas characterized by two degrees of freedom ($n=2$) and the 
fundamental equation $U(S,V)= [\exp(S/\kappa)/V]^{2/3}$, where $\kappa=$ const.  \cite{callen}.
It is convenient to use
the entropy representation in which the first law reads 
$ dS= (1/T) dU + (P/T) d V$, and the equilibrium conditions
are $1/T = \partial S/\partial U$ and $P/T = 
\partial S /\partial V$. Consequently, for the phase space we can use the coordinates 
$ Z^A = \left\{S,U,V,{1}/{T},{P}/{T}\right\}$  and
the metric (\ref{ginv2}) takes the form
\be
\label{gups}
G = \left(dS -\frac{1}{T} d U - \frac{P}{T} dV\right)^2
+ \Lambda \left[\left(\frac{U}{T}\right)^{2k+1} dU d\left(\frac{1}{T}\right) 
+\left(\frac{VP}{T}\right)^{2k+1} dV d\left(\frac{P}{T}\right)\right] \ .
\ee
Moreover, the  metric for the space of equilibrium can be derived
from Eq.(\ref{gdowninv}):  
\bea
\label{gdowns}
g= \Lambda &\Bigg\{&\left(U\frac{\partial S}{\partial U}\right)^{2k+1}\frac{\partial^2 S}
{\partial U^2} dU^2
+  \left(V\frac{\partial S}{\partial V}\right)^{2k+1}
\frac{\partial^2 S}{\partial V^2} dV^2 \nonumber\\
& +&\left[ \left(U\frac{\partial S}{\partial U}\right)^{2k+1}
+ \left(V\frac{\partial S}{\partial V}\right)^{2k+1} \right] 
\frac{\partial^2 S}{\partial U \partial V} dU dV \ \Bigg\} \ .
\eea
This form of the thermodynamic metric is valid for any 
 system with two degrees of freedom represented by the extensive
variables $U$ and $V$. To completely determine the metric it is only necessary to specify the fundamental equation 
$S=S(U,V)$. For an ideal gas $S(U,V) = \frac{3\kappa}{2}\ln U + \kappa\ln V$, and the metric becomes 
\be
\label{gdownig}
g= - \kappa^{2k+2}\Lambda\left[ \left(\frac{3}{2}\right)^{2k+2} \frac{dU^2}{U^2}
+ \frac{dV^2}{V^2}\right]\ .
\ee
All the geometrothermodynamical information about the ideal gas must be contained 
in the metrics (\ref{gups}) and (\ref{gdownig}). First, we must show that 
the subspace of equilibrium $({\cal E},g)$ determines an extremal hypersurface
in the phase manifold $({\cal T}, G)$. Introducing (\ref{gups}) and (\ref{gdownig})  into the 
Nambu-Goto equations (\ref{meqg}), the system reduces to 
\bea
\label{condig1}
\frac{\partial\Lambda}{\partial U} 
+ \frac{3\kappa}{2U^2} \frac{\partial \Lambda}{\partial Z^3}
+ 2(k+1)\frac{\Lambda}{U} & = & 0\ , \\
\frac{\partial\Lambda}{\partial V}
+ \frac{\kappa}{V^2} \frac{\partial \Lambda}{\partial Z^4}
+ 2(k+1)\frac{\Lambda}{V} & = & 0\ .
\label{condig2}
\eea
If we choose $\Lambda=const$ and $k=-1$, we obtain 
a particular solution which is probably the simplest one. This shows that the geometry
of the ideal gas is a solution to the differential equations of GTD and, consequently,
determines an extremal hypersurface of the thermodynamic phase space. 

We now investigate the geometry of the space of equilibrium states of the ideal gas.
As can be seen from Eq.(\ref{gdownig}), the geometry is described by a 2-dimensional
conformally flat metric. If we calculate the curvature scalar $R$, and replace in the result the 
the differential conditions (\ref{condig1})
and (\ref{condig2}), we obtain
\bea
R\propto &&
3V^2\left\{2U^2\Lambda \frac{\partial^2\Lambda}{\partial U \partial Z^3}
+ \frac{\partial\Lambda}{ \partial Z^3} 
\left[3\kappa \frac{\partial\Lambda}{ \partial Z^3} + 2(k+1) U \Lambda\right]\right\}
\nonumber \\
&& +\,4\left(\frac{3}{2}\right)^{2k+2}U^2
\left\{V^2\Lambda \frac{\partial^2\Lambda}{\partial V \partial Z^4}
+ \frac{\partial\Lambda}{ \partial Z^4} 
\left[\kappa \frac{\partial\Lambda}{ \partial Z^4} + 2(k+1) V \Lambda\right]\right\}\ .
\eea
We see that it is  
possible to choose the conformal factor $\Lambda$ such that $R=0$. For instance, 
the choice $\Lambda=const$ is a particular solution which also satisfies the 
differential conditions (\ref{condig1}) and (\ref{condig2}) for $k=-1$. Consequently,
we have shown that the ideal gas can be represented by a flat metric in the space 
of equilibrium states. This result agrees with our intuitive expectation that 
a thermodynamic metric with zero curvature should describe a system in which no
thermodynamic interaction is present. The freedom involved in the choice of the 
function $\Lambda$ is associated to the well-known fact that any 2--dimensional space
is conformally flat. 

We now investigate 
the geodesic equations (\ref{geo1}).  For concreteness we take the values  
$\Lambda=-1$ and $k=-1$. Then, the metric 
takes the form $g=dU^2/U^2 + dV^2/V^2$ which can be put in the obvious flat form 
\be
g = d\xi^2 + d\eta^2
\ee
by means of the transformation $\xi= \ln U ,\eta= \ln V $,  where for simplicity 
we set the additive constants of integration such that $\xi,\eta\geq 0$. 
The solutions of the geodesic equations are represented by straight lines,
$\xi=c_1\eta + c_0$, with constants $c_0$ and $c_1$.
In this representation, the entropy becomes a simple linear function of the coordinates,
$S=(3\kappa/2)\xi + \kappa \eta$. Since each point on the
$\xi\eta-$plane can represent an equilibrium state, the geodesics should connect
those states which are allowed by the laws of thermodynamics. For instance, 
consider all geodesics with initial state $\xi=0$ and $\eta=0$. Then, any straight 
line pointing outwards of the initial state and contained inside 
the allowed positive quadrant connect states with increasing entropy. 
A quasi-static process connecting
states in the inverse direction is not allowed by the second law. 
Consequently, the affine parameter $\tau$ along the geodesics can actually be 
interpreted as a time parameter and the direction of the geodesics indicates 
the ``direction of time". 
A detailed analysis of these geodesics is presented elsewhere \cite{qsv08}.

\subsection{The van der Waals gas}
\label{sec:vdw}
A more realistic model of a gas, which takes into account the size of the particles 
and a pairwise attractive force between the particles of the gas, is based upon 
the van der Waals fundamental equation 
\be
\label{feqvdw}
S = \frac{3 \kappa}{2} \ln\left(U+ \frac{a}{V}\right) + \kappa \ln(V - b) \ ,
\ee
where $a$ and $b$ are constants. 
The metric of the manifold ${\cal T}$ is as before (\ref{gups}). For simplicity we consider 
the particular case with $k=-1$. Then, 
introducing the fundamental equation (\ref{feqvdw}) into the metric 
(\ref{gdowns}), the metric of the manifold ${\cal E}$
reads
\bea
\label{gdownvdw}
g = \frac{\Lambda}{U(U+a/V)} \Bigg[ && - dU^2 + \frac{U}{V^3} 
\frac{a(a+2UV)(3b^2-6bV+V^2)-2U^2V^4}{(V-b)(3ab-aV+2UV^2)}\, dV^2 \nonumber \\
&&+ \frac{a}{V^2}\frac{3ab-aV-3bUV+5UV^2}{3ab-aV+2UV^2} dU dV\Bigg] \ .
\eea
The curvature of this thermodynamic metric is in general nonzero, 
reflecting the fact that the thermodynamic interaction of the 
van der Waals gas is nontrivial. 
The differential equations (\ref{meqg}) can be computed for this case
by using the phase space metric (\ref{gups}), with $k=-1$, and the metric 
(\ref{gdownvdw}) for the space of equilibrium. It turns out that they 
reduce to only two first--order partial differential equations
\bea
 \label{condvdw1}
\frac{\partial\Lambda}{\partial U} 
+ F_3 \frac{\partial \Lambda}{\partial Z^3}
+ F_4 \frac{\partial \Lambda}{\partial Z^4}
+ F_0 \Lambda & = & 0\ , \\
\frac{\partial\Lambda}{\partial V}
+ G_3 \frac{\partial \Lambda}{\partial Z^3}
+ G_4 \frac{\partial \Lambda}{\partial Z^4}
+ G_0 \Lambda & = & 0\ , 
\label{condvdw2}
\eea
where $F_0, F_3, F_4,G_0,G_3$, and $G_4$ are fixed rational functions of $U$ and $V$. 
Because of the arbitrariness of the conformal factor $\Lambda$, it is always possible to find
solutions to the above system. 
We conclude that there exists a family of nonflat 
thermodynamic metrics that determines an extremal hypersurface 
in the phase space, and can be used to describe the geometry of
the van der Waals gas. 
The corresponding geodesic equations  are highly nontrivial and require a detailed 
numerical analysis which is beyond the scope of the present work.

\subsection{New solutions}
\label{sec:new}
The above applications of GTD show that from a given fundamental equation one can find the 
corresponding geometric representation. It is also possible to handle the inverse problem, i.e.,
we can find fundamental equations which are compatible with the geometric structures of GTD
and correspond to thermodynamic systems. 
Consider a very simple generalization of the ideal gas  
\be
\label{genig}
S(U,V) = \frac{3\kappa}{2}\ln U + \kappa c \ln V\ ,
\ee
where $c$ is a constant. Although this seems to be a trivial generalization, we will 
see that it can contain interesting thermodynamic systems. 
As before, we choose the thermodynamic metrics in ${\cal T}$ and ${\cal E}$ as 
in Eqs.(\ref{gups}) and (\ref{gdowns}). The Nambu-Goto equations 
(\ref{meqg}) are identically satisfied if we choose $\Lambda=-1$ and 
$k=-1$. Then, the space of states turns out to
be flat and, according to our interpretation of thermodynamic curvature, 
the system is characterized by the absence of thermodynamic interaction. 
The geometric analysis of the manifold ${\cal E}$ is similar to that carried out 
in Section \ref{sec:ig}.
To interpret this solution it is convenient to use the
energy representation in which the fundamental equation is 
$
U(S,V)=\frac{e^{2S/3\kappa}}{V^{2c/3}}$. 
The conditions for equilibrium (\ref{condte}) and Euler's identity
(\ref{euler}) lead to 
\be
\frac{\partial U}{\partial S} = T = \frac{2}{3\kappa} U\ ,\quad
\frac{\partial U}{\partial V} = -P = -\frac{2c}{3} \frac{U}{V}\ ,\quad
S=\frac{3\beta\kappa}{2}+c\kappa \ ,
\ee
where $\beta$ is the degree of homogeneity of the thermodynamic potential. Hence,
\be
U=\frac{3\kappa}{2} T\ ,\quad PV = \kappa c T \ ,
\ee
are the equations of state. The internal energy of the system coincides with that of an ideal gas, and 
the only difference  appears in the behavior of the pressure $P$ which can be 
controlled by the constant $c$. If we define the energy density $\rho=U/V$, then 
from the above equations we obtain the ``barotropic" equation of state $P=(2c/3)\rho$.
For the particular choice $c=-3/2$, we obtain $P+\rho=0$ with a negative value 
of the pressure. Consequently, the fundamental equation (\ref{genig}) describes 
a system with no thermodynamic interaction and negative pressure. 
This behavior resembles that of the dark energy which is responsible for the 
recently observed acceleration of the Universe. A more detailed analysis will
be necessary to determine if the above thermodynamic system can be used as a
realistic model for dark energy.  

The above example can be generalized to include a complete family of noninterac\-ting
thermodynamic systems. In fact, if we consider a system with $n$ degrees of 
freedom and the separable 
fundamental equation 
\be
S(E^1,\cdots,E^n) = S_1(E^1) + \cdots + S_n(E^n)\ ,
\ee  
where $S_1, \cdots, S_2$ are arbitrary smooth functions, 
we obtain  from Eq.(\ref{gdowninv}), with $\Lambda= $const,
 a diagonal thermodynamic metric of the form 
$g=g_{11}(E^1) (dE^1)^2 +\cdots +  g_{nn}(E^n) (dE^n)^2$. The curvature of this 
metric vanishes as can be seen by applying the coordinate transformation 
$g_{aa}(E^a) dE^a = d X^a$ (no summation over repeated indices)  
which transforms the metric into
the  flat form $g=\delta_{ab} dX^a d X^b$.  In this family of
thermodynamically noninteracting systems one can include, for instance,
all known multicomponent generalizations of the ideal gas. 

Turning back to the case of systems with two degrees of freedom, we mention that
it is possible to find complete classes of solutions of the form 
\be 
S= S_0 U^\alpha V^\beta\ \qquad {\rm or}  \qquad 
S=S_0 \ln (U^\alpha + cV^\beta)\ ,
\ee
where $S_0$, $\alpha$,
$\beta$, and $c$ are arbitrary real constants. It turns out that in all these cases,
one can choose $\Lambda$ and $k$ in such a way 
that the resulting thermodynamic metric (\ref{gdowns}) is curved and satisfies the
Nambu-Goto  equations (\ref{meqg}). This means that the above fundamental
equations can, in principle, represent nontrivial thermodynamically interacting 
systems. It is not difficult to find exact solutions to the 
Nambu-Goto equations which are also compatible with the metric structure (\ref{gups})
of the phase manifold and, consequently, with the thermodynamic metric of the manifold
of equilibrium states. 
Nevertheless, a more detailed analysis will be necessary in order to establish
the physical significance of the solutions obtained in this manner.

\section{Discussion and conclusions}
\label{sec:con}

Geometrothermodynamics (GTD) is a formalism that has been developed recently to describe 
ordinary thermodynamics by using differential geometry.  To this end, two known structures were joint together in a consistent 
manner: The natural contact structure of the thermodynamic phase space ${\cal T}$
and the metric structure of the space of equilibrium states ${\cal E}$. 
In GTD, one introduces a Legendre invariant metric $G$ in ${\cal T}$ which  
induces a thermodynamic metric $g$ in ${\cal E}$ by means of the pullback $\varphi^*$
associated to the map $\varphi: {\cal E} \mapsto {\cal T}$. This additional 
construction confers to ${\cal T}$ and ${\cal E}$ the geometric structure of 
Riemannian manifolds. 
In this work we 
established that $\varphi$ can be handled as a harmonic map that allows us to introduce 
a Polyakov-like action in ${\cal E}$. The variation of the harmonic map action 
leads to a set of second--order differential equations which can be identified as
the Nambu-Goto motion equations. 
This is the main result that allows us to interpret
thermodynamic systems as classical bosonic strings. 
In GTD, thermodynamic systems are characterized by a specific metric $g$ 
which determines the properties of ${\cal E}$. 
Therefore, if $g$ satisfies the Nambu-Goto 
equations, we can conclude that it describes an $n-$dimensional membrane that ``lives" 
in the background manifold ${\cal T}$. If we 
demand Legendre invariance of $G$, the background
${\cal T}$ turns out to be curved in general. So 
the explicit case of a thermodynamic system with two degrees of freedom and thermodynamic 
metric $g$ resembles the dynamics of a string moving on a nonflat background $G$. 
The analogy, however, is only at the mathematical level. In fact, in string theory 
the metrics $g$ and $G$ must be Lorentzian metrics in order to incorporate into the
theory a relativistic dynamical behavior with a genuine time parameter. In GTD the metrics
are not necessarily Lorentzian and there is no explicit time parameter 
so that we cannot really talk about dynamics. This is in agreement with our 
intuitive understanding of ordinary 
thermodynamics of equilibrium states in which, strictly speaking, 
there is no dynamics at all and we are in fact handling with thermostatics. 
Non-equilibrium thermodynamics is a different subject that cannot be incorporated
in a straightforward manner in GTD as formulated here. A generalization of the 
geometric structures considered in this work will be necessary in order to analyze 
more general scenarios in which non-equilibrium states cannot be neglected. This is 
a task for future investigations. 

Starting from the most general Legendre invariant metric in ${\cal T}$ we were able to
show that the ideal gas and the van der Waals gas are concrete examples of 2-dimensional
extremal hypersurfaces ${\cal E}$ embedded in a 5-dimensional curved manifold ${\cal T}$.
In the case of an ideal gas, the geometry of ${\cal E}$ is flat, whereas for the van der 
Waals gas the manifold ${\cal E}$ is curved. This reinforces the interpretation 
of the thermodynamic curvature as a measure of thermodynamic interaction.  
Our formalism is such that one only needs to postulate an arbitrary fundamental 
equation to derive the thermodynamic metric $g$ of ${\cal E}$. 
One can then derive from the Nambu-Goto  equations 
the conditions that the fundamental equation and $g$ must satisfy in order to 
correspond to an extremal hypersurface of ${\cal T}$. In this manner, we obtained 
simple generalizations of the ideal gas with vanishing thermodynamic curvature. 
It is also possible  to derive new solutions with nonvanishing
thermodynamic curvature.  

We used the concept of thermodynamic length in the manifold of equilibrium 
states ${\cal E}$ as a quantity representing the geometric distance between two different
states. 
By demanding that the thermodynamic length be an extremal, we obtained that the
geodesic equations for $g$ must be satisfied. Not all the solutions of the geodesic
equations need to be physically realizable since they could connect
 equilibrium states that are not compatible with the laws of thermodynamics.
We interpret geodesics which connect thermodynamically meaningful states as 
representing a dense succession of quasi-static states. Moreover, we showed that in an 
ideal gas 
the affine parameter along a geodesic can be used to label each of the equilibrium 
states which can be reached in a specific quasi-static thermodynamic process. 
The affine parameter can then be chosen in such a way that it increases as the 
entropy of a quasi-static process increases. Then, a suitable selection allows us 
to interpret the affine parameter as a ``time" parameter with a specific direction 
which coincides with the direction of entropy increase. 
The geodesic equations for the more general van der Waals gas cannot be solved
analytically. It will be necessary to perform a detailed numerical study of 
these equations in order to  corroborate the physical significance 
of the geodesics.

\section*{Acknowledgements}
This work was supported in part by Conacyt, Mexico, Grants No. 48601 and 165357. One of us (HQ) would like
to thank ICRANet for support.

\end{document}